\documentstyle[aps,12pt]{revtex}

\begin{document}
\title{The orbital moment in CoO}
\author{R. J. Radwanski}
\address{Center for Solid State Physics, S$^{nt}$ Filip\\
5,31-150Krakow,Poland.}
\author{Z. Ropka}
\address{Center for Solid State Physics, S$^{nt}$ Filip\\
5,31-150Krakow,Poland.\\
email: sfradwan@cyf-kr.edu.pl, http://css-physics.edu.pl}
\maketitle

\begin{abstract}
The orbital and spin moment of the Co$^{2+}$ ion in CoO has been calculated
within the quasi-atomic approach with taking into account the intra-atomic
spin-orbit coupling. The orbital moment of 1.38 $\mu _{B}$ amounts at 0 K,
in the magnetically-ordered state, to more than 34\% of the total moment
(4.02 $\mu _{B}$) and yields the L/S ratio of 1.04, close to the
experimental value.


\end{abstract}

\date{(14.05.2002)}

CoO attracts a large attention of the magnetic community by more than 50
years. Despite of its simplicity (two atoms, $NaCl$ structure, well-defined
antiferromagnetism (AF) with T$_{N}$ of 290 K) and enormous theoretical and
experimental works the consistent description of its properties, reconciling
its insulating state with the unfilled 3$d$ band is still not reached \cite%
{1,2,3,4,5,6,7,8,9}.

The aim of this short Letter is to report the calculations of the magnetic
moment of CoO. The direct motivation was a just published paper of Ref. \cite%
{9}. In our approach we attribute the moment of CoO to the Co$^{2+}$ ions.
We have calculated the moment of the Co$^{2+}$ ion in the CoO$_{6}$
octahedral complex, its spin and orbital parts, and the orbital moment as
large as 1.38 $\mu _{B}$ at 0 K has been found. The approach used can be
called the quasi-atomic approach \cite{10,11} as the starting point for the
description of a solid is the consideration of the atomic-like low-energy
electronic structure of the constituting atoms/ions, in the present case of
the Co$^{2+}$ ions.

We have treated the 7 outer electrons of the Co$^{2+}$ ion as forming the
highly-correlated electron system 3$d^{7}$. The correlations among electrons
in the unfilled 3$d$ shell are approximated by two Hund's rules, that yield
the ground-term quantum numbers $S$=3/2 and $L$=3, i.e. the ground term $%
^{3}F$ \cite{12,13}. Such the localized highly-correlated electron system
interacts in a solid with the charge and spin surroundings. The charge
surrounding has the octahedral symmetry owing to the $NaCl$-type of
structure of CoO. We take into account the small tetragonal distortion as is
experimentally observed \cite{8}. The tetragonal distortion is important for
the detailed formation of the AF structure and influences the spin and
orbital moments but the most essential physical interaction is the
intra-atomic spin-orbit coupling. Our Hamiltonian for CoO consists of two
terms: the single-ion-like term $H_{d}$ of the 3$d^{7}$ system and the $d$-$%
d $ intersite spin-dependent term, important for the formation of the
magnetic state. The calculations follow those, that we have performed for
the description of FeBr$_{2}$ and NiO \cite{10,11,14}. For the quasi-atomic
single-ion-like Hamiltonian of the 3$d^{7}$ system we take into account the
crystal-field interactions of the octahedral symmetry (the octahedral CEF
parameter B$_{4}$=-40 K), the spin-orbit interactions with the spin-orbit
coupling $\lambda _{s-o}$=-260 K and the tetragonal distortion approximated
by the term B$_{2}^{0}$=-10 K. The calculated single-ion states under the
octahedral crystal field and the spin-orbit coupling (the CoO$_{6}$ complex)
are presented in Fig. 1.

The ground-state doublet characterized by the total moment of $\pm $2.21 $%
\mu _{B}$ is built up from the orbital and spin moments of $\pm $0.70 and $%
\pm $1.84 $\mu _{B}$, respectively. It, however, fully cancels in the
paramagnetic state and reveals itself only in the presence of the magnetic
field, external or internal in case of the magnetically-ordered state, that
polarizes two doublet states. This doublet is the Kramers doublet. The
intersite spin-dependent interactions cause the (antiferro-)magnetic
ordering. They have been considered in the mean-field approximation with the
molecular-field coefficient $n$ acting between magnetic moments as -40T/ $%
\mu _{B}$. It means that the Co moment in the magnetic state at 0 K
experiences the molecular field of 161 T. It causes the spin-like gap of 40
meV.

The calculated value of the magnetic moment at 0 K in the
magnetically-ordered state amounts to 4.02 $\mu _{B}$. It is built up from
the spin moment of 2.64 $\mu _{B}$ ($S_{z}$=1.32) and the orbital moment of
1.38 $\mu _{B}$. The increase of the orbital moment in comparison to the
paramagnetic state is caused by the polarization of the ground-state
eigenfunction by the molecular magnetic field. The orbital moment is quite
substantial being more than 34\% of the total moment. Our theoretical
outcome, revealing the substantial orbital moment is in nice agreement with
the very recent experimental result of 3.98$\pm $0.06 $\mu _{B}$ for the Co
moment \cite{8}. The magnetic x-ray experiment has revealed the $L$/$S$
ratio of 0.95 \cite{15}- the calculated by us values lead to $L$/$S$ ratio,
in fact their $z$ components, of 1.04. In NiO we have calculated the $L$/$S$
ratio as 0.54 at 0 K \cite{14}.

The calculated temperature dependence of the total moment, together with the
orbital and spin moments, is shown in Fig. 3. These moments disappear above T%
$_{N}$ - in the paramagnetic region the derived moment is the effective
moment, that bears the information about $J^{2}$ or $S^{2}$.

We would like to point out that the evaluation of the orbital moment is
possible provided the spin-orbit coupling is taken into account. It confirms
the importance of the spin-orbit coupling for the description of the 3$d$%
-ion compounds. The present model allows, apart of the ordered moment and
its spin and orbital components to calculate many physically important
properties like temperature dependence of the magnetic susceptibility,
temperature dependence of the heat capacity, the spectroscopic $g$ factor,
the fine electronic structure in the energy window below 4 eV with at least
28 localized states.

Finally, we would like to point out that our approach should not be
considered as the treatment of an isolated ion only - we consider the Co$%
^{2+}$ ion in the oxygen octahedron. The physical relevance of our
discussion to CoO is obvious - the NaCl structure is built up from the edge
sharing Co$^{2+}$ octahedra and the similar electronic structure occurs at
each Co site.

In conclusion, the orbital and spin moment of the Co$^{2+}$ ion in CoO has
been calculated within the quasi-atomic approach taking into account the
intra-atomic spin-orbit interactions. The orbital moment of 1.38 $\mu _{B}$
amounts at 0 K, in the magnetically-ordered state, to more than 34\% of the
total moment (4.02 $\mu _{B}$). Our atomic-like approach provides the
discrete energy states for 3$d$-electrons in CoO and calculates, apart of
the spin moment, the orbital moment. This orbital moment is completely
ignored in most of modern solid-state physics theories. Our studies indicate
that it is the highest time in solid-state physics to ''unquench'' the
orbital moment in the description of 3$d$-atom containing compounds, the
more that it becomes visible in recent advanced experiments.

{\bf Fig. 1.} The electronic structure of the Co$^{2+}$ ion in CoO. a) the
free-ion $^{4}F$ term, b) the effect of the octahedral crystal field, c) the
electronic structure resulting from the octahedral crystal field and the
intra-atomic spin-orbit interactions.

Fig. 2. The splitting of the ground-state Kramers doublet in the
magnetically-ordered state of CoO.

Fig. 3. Temperature dependence of the Co$^{2+}$-ion moment in CoO. At 0 K
the total moment of 4.02 $\mu _{B}$ is built up from the orbital and spin
moment of 1.38 and 2.64 $\mu _{B}$, respectively. The calculations have been
performed for the quasi-atomic parameters with the octahedral crystal-field
parameter B$_{4}$= -40 K, the spin-orbit coupling constant $\lambda _{s-o}$=
-260 K, the intersite spin-dependent interactions given by the
molecular-field coefficient $n$ = -40 T/ $\mu _{B}$ and the tetragonal
distortion parameter B$_{2}^{0}$ of -10 K.

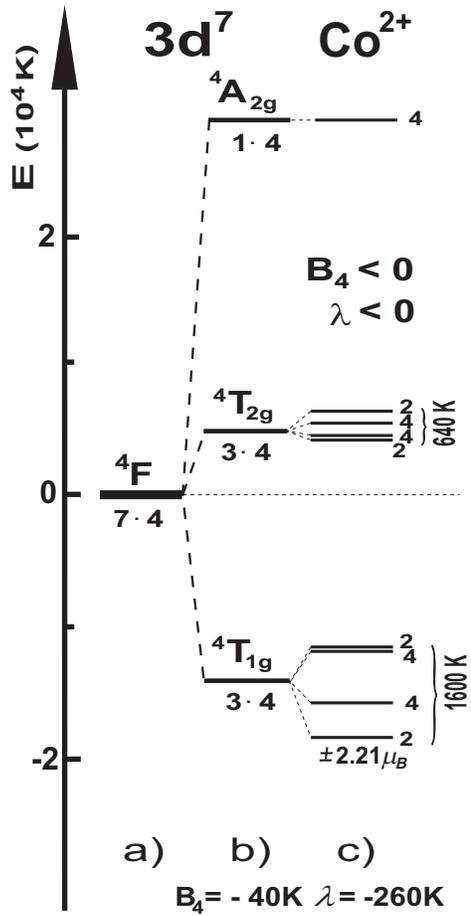
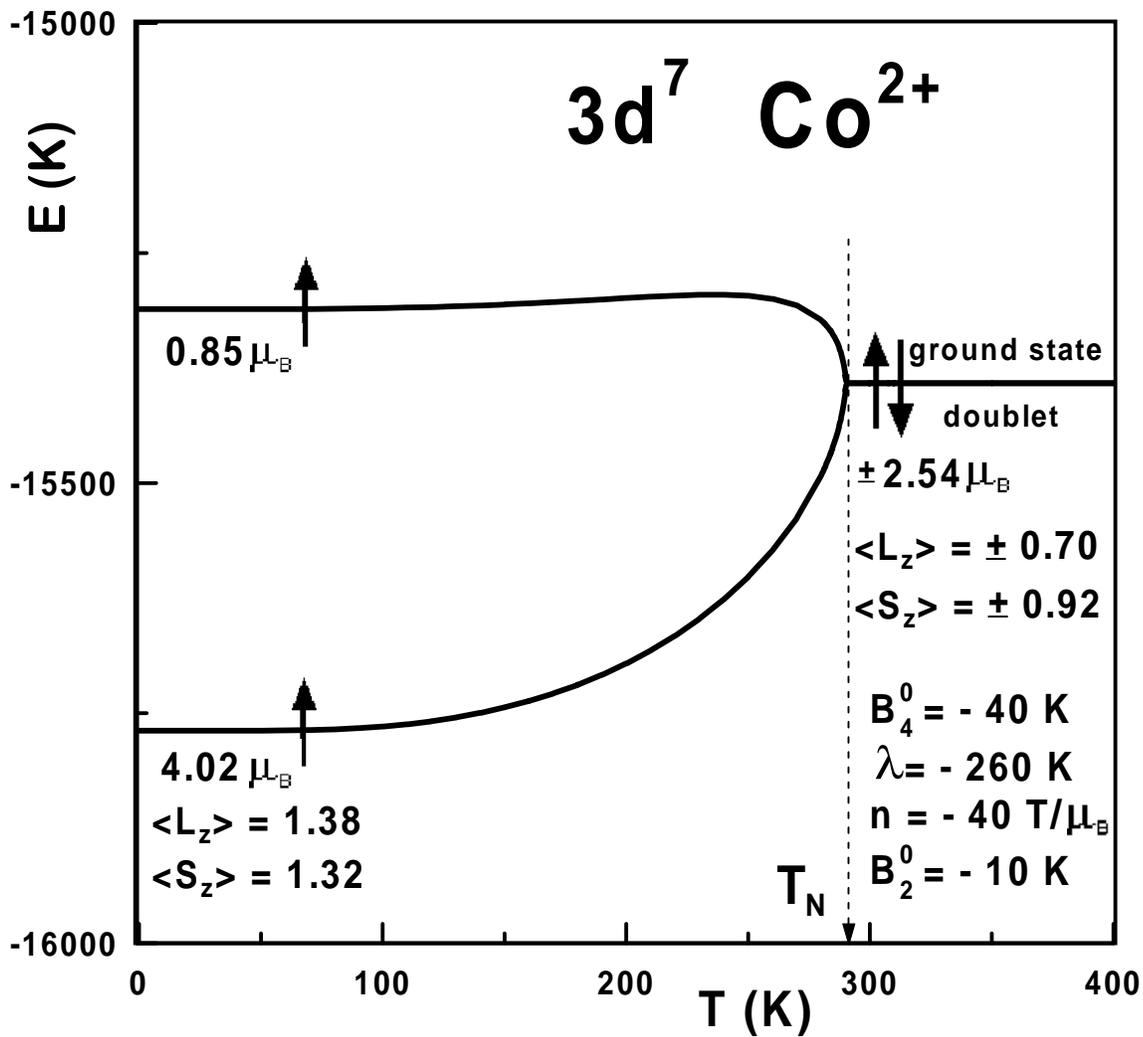
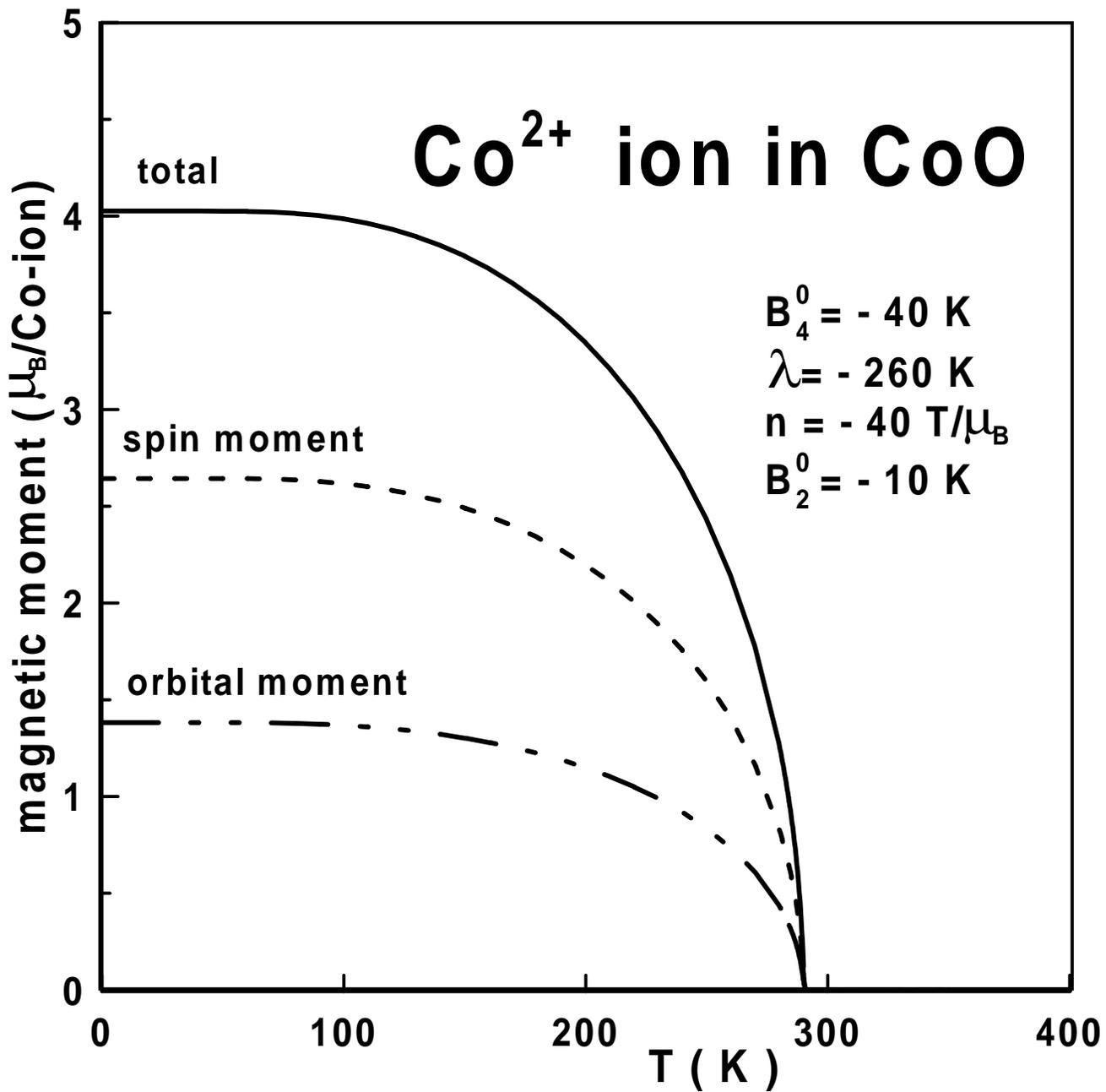
\end{document}